\documentclass[useAMS, usenatbib, usegraphicx]{mn2e}
\usepackage{times}

\newcommand{\V}{{\rm {\scriptsize V}}}
\newcommand{\I}{{\rm {\scriptsize I}}}

\title[The ejecta of V1494~Aql]{Six months of mass outflow and
 inclined rings in the ejecta of V1494~Aql}

\author[S.~P.~S.~Eyres et al.]{S.~P.~S.~Eyres$^1$, I. Heywood$^2$,
T.~J.~O'Brien$^2$, R.~J.~Ivison,$^3$ T.~W.~B.~Muxlow$^2$, \newauthor V. G. Elkin$^1$\\
$^1$Centre for Astrophysics, University of Central Lancashire,
Preston, PR1 2HE, UK\\
$^2$Jodrell Bank Observatory, University of Manchester,
Macclesfield, Cheshire, SK11 9DL, UK\\
$^3$Astronomy Technology Centre, Royal Observatory,
Blackford Hill, Edinburgh, EH9 3HJ, UK\\
}

\date{Received; in original form}

\begin{document}
\label{firstpage}
\maketitle

\begin{abstract}
V1494~Aql was a very fast nova which reached a visual maximum of
m$_{\rm v}\simeq$4.0 by the end of 1999~December~3. We report
observations from 4 to 284~days after discovery, including sub--mm--
and cm--band fluxes, a single MERLIN image, and optical spectroscopy
in the 410 to 700~nm range.  The extent of the radio continuum
emission is consistent with a recent lower distance estimate of
1.6~kpc. We conclude that the optical and radio emission arises from
the same expanding ejecta. We show that these observations are not
consistent with simple kinematical spherical shell models used in the
past to explain the rise and fall of the radio flux density in these
objects. The resolved remnant structure is consistent with an inclined
ring of enhanced density within the ejecta. Optical spectroscopy
indicates likely continued mass ejection for over 195~days, with the
material becoming optically thin in the visible sometime between 195
and 285~days after outburst.
\end{abstract}


\section{Introduction}
\label{sec-intro}

When a classical nova undergoes an eruption, the initial visual
brightening is accompanied by the ejection of perhaps 10$^{-5}$ to
10$^{-4}$~M$_\odot$ of matter.  This expands away from the central
binary, and as the visual light starts to decline the increasing solid
angle of the ionised ejecta leads to steadily rising, optically thick,
thermal radio emission.  As the remnant expands the emission becomes
optically thin at ever longer wavelengths, and fades below
detectability.  This process is understood in general terms, but the
development of structure in the remnant observed a few months after
outburst is still unclear. Asymmetric ejection, intra--ejecta
interaction and the influence of the binary companion all have a role
to play.

Much still needs to be done observationally to constrain these models.
Examining the very early development of the remnant is one means by
which this is currently taking place.  Radio observations hold out the
possibility of imaging the remnant structure at very early times,
particularly with MERLIN providing 50~mas resolution.  Observations to
date \citep{Eyres96, Eyres00} demonstrate that the clumpy,
non--spheroidal structures seen at late times \citep[e.g.][]{Gill98,
Gill00} appear very early, and also develop rapidly.  For V1974~Cyg,
in which the shell structure was resolved by MERLIN only eighty days
after outburst \citep{Pavelin93, Eyres96}, \citet{Ivison93} found the
sub--mm to mm data early in the outburst were inconsistent with
kinematical models used to fit radio data at later times. In many
novae optical spectroscopy is consistent with axisymmetric structures,
with features due to density enhancements in polar caps, tropical
rings and equatorial bands evident to different degrees in different
novae \citep[e.g.][]{Gill99}.  In V705~Cas the radio images show the
structures developing asymmetrically while the optical spectra are
consistent with an axisymmetric density distribution
\citep{Eyres00}. V723~Cas showed a similar development to V1974~Cyg
\citep{Heywood04}.
Observing further novae over a range of speed classes \citep[see][for
a definition of these classes]{Payne-Gaposchkin64} and hence ejecta
velocities is important in constraining revised models over the
physical parameter space.

V1494~Aql was first detected at m$_{\rm v}\simeq$6.0 on
1999~December~1 \citep{Pereira99}, reaching visual maximum of m$_{\rm
v}\simeq$4.0 by the end of 1999~December~3 \citep{Rao99, Liller99}. It
dropped to fainter than m$_{\rm v}\simeq$6.0 by 1999~December~10
\citep{Pontefract99} and m$_{\rm v}\simeq$7.0 by 1999~December~19
\citep{Keen99}. \citet{Kiss00} find t$_2$~=~6.6$\pm$0.5~days and
t$_3$~=~16$\pm$0.5~days, making it a very fast nova. Optical
spectroscopy has identified velocity components at up to
1700~km~s$^{-1}$ \citep{Fujii99} while IR spectroscopy shows
velocities up to 2900~km~s$^{-1}$ \citep{Rudy01}, consistent with this
classification.  \citet{Iijima03} concluded that mass ejection
continued for over 150~days, contradicting a more general conclusion
that ejection in novae ceases after a few days
\citep[e.g.][]{Hjellming96}. In this work we present the results of
observations of V1494~Aql starting on 1999~December~7, within 5~days
of its discovery in outburst and interpret them in the context of
standard models of mm-cm radio emission from novae.

\section{Observations}
\label{sec-observations}

We report observations in the radio, sub--mm and optical bands using
 the Multi--Element Radio--Linked Interferometer Network (MERLIN), the
 Very Large Array (VLA), the James Clerk Maxwell Telescope (JCMT), and
 the 1--m Zeiss telescope at the Special Astrophysical Observatory
 (SAO) of the Russian Academy of Science.  Observing dates range from
 4 to 288~days after visual maximum (1999~December~3, taken as day~0
 hereafter).

\subsection{JCMT}
\label{ssec-JCMT_obs}

Initial observations were made with the JCMT on 1999~December~7 and 8
(days~4 and 5), at 450$\umu$m and 850~$\umu$m using SCUBA in
photometry mode \citep{Ivison98, Holland99} in good weather (optical
depth $\tau \sim 0.07$ at 225~GHz).  Calibration was made against the
JCMT primary calibrator Uranus, with skydips to measure opacity at
both observing wavelengths.

\subsection{Radio observations}
\label{ssec-MERLIN_obs}
\label{ssec-VLA_obs}

MERLIN Target of Opportunity observations were triggered and took
place on 1999~December~30 (day~27) at 18~cm.  The observations were
made in phase--referenced mode, using 1919+063 as the phase--reference
source.  The flux density scale was set by comparison of the
unresolved source OQ208 with the resolved flux density calibrator
3C286.  On the shortest MERLIN spacings at 1658~MHz 3C286 has a flux
density of 13.639~Jy. OQ208 by comparison has a flux density 1.125~Jy,
giving the calibrated flux density of 1919+063 to be 0.164~Jy. The
data were mapped in wide--field mode to deal with a bright confusing
source in the side--lobes of the primary beams of the antennas.  A
later MERLIN observation was made on 2000~April~17 (day 136) at 6~cm, using
1929+050 as a phase--reference source with a calibrated flux density
of 0.27~Jy.  The data for V1494~Aql were phase calibrated using the
solutions for the phase--reference source. 

Ad Hoc VLA data were obtained on 2000~June~7 (day~187) in C--array.
Observations were made at 20, 6.1 and 3.5~cm, with a bandwidth of
50~MHz in each case.  The primary calibrator was 1331+305 (=~3C286),
which was used to calibrate the secondary calibrator 1922+155, for
which flux densities of 0.61$\pm$0.04, 0.632$\pm$0.007 and
0.63$\pm$0.01~Jy were determined in each band respectively.  This was
used to flux-- and phase--calibrate the observations of V1494~Aql.

\subsection{Optical spectroscopy}
\label{ssec-optical_spec}

Observations were made with the Universal Astronomical Grating
Spectrograph (UAGS) on the 1--m Zeiss telescope at the SAO on
2000~June~14 \& 15 (days~194 \& 195) and September~9 \& 14 (day~281 \&
286). The wavelength calibration was carried out using a neon arc with
bespoke software for the reduction of data from this instrument
\citep{Vlasyuk93}. The observations were made with two different
gratings, yielding spectral resolutions of around 1000 (June) and 2000
(September), giving a velocity resolution of 300 and 150~km~s$^{-1}$
respectively. The data are not flux calibrated.

These observations were made when V1494~Aql was still bright. We see
no evidence for contamination of the spectra by features from the
nearby G--type star, as seen at later times by \citet{Kiss04}.

\section{Results}
\label{sec-results}

\begin{figure}
\includegraphics[angle=270, width=8.5cm]{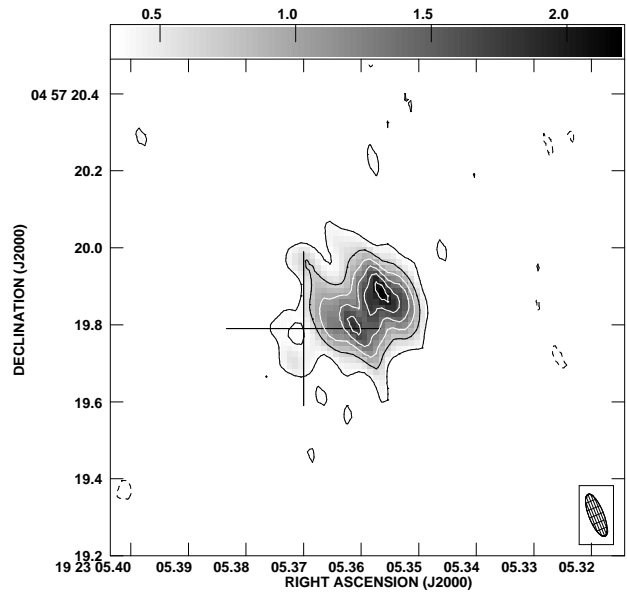}

\caption{MERLIN image of the remnant of V1494~Aql from 2000~April~17,
day~136. The optical position of \citet{Kiss04} is marked as a cross, with the
size indicating the uncertainty. The contours are $-$3, 3, 6, 9, 12,
15, 18$\times$0.114~mJy~beam$^{-1}$ and greyscale runs from 0.4 to
2.1~mJy~beam$^{-1}$.  The peak is 2.10$\pm$0.11~mJy~beam$^{-1}$.}
\label{fig-MERLIN}
\end{figure}

On December 7 \& 8 (days 4 \& 5 after maximum), the flux densities,
taking an integration time weighted average over both nights, were
22.5$\pm$1.6~mJy at 850~$\umu$m and 114$\pm$28~mJy at 450~$\umu$m,
with marginal evidence of an increase between the two nights at
850~$\umu$m (rise of 5.9~mJy compared with a combined uncertainty in
the difference of 5.5~mJy). By the time of the MERLIN observations on
day~27, the total flux density was 0.50$\pm$0.07~mJy at 18~cm.  The
emission was unresolved at this time, as expected.  By day~136 the
remnant was partially resolved with MERLIN at 6~cm , and the total
detected flux density was found to be 25.0$\pm$1.2~mJy (see
Fig.~\ref{fig-MERLIN}). The emission detected with MERLIN is
consistent with the optical position from \citet{Kiss04} of 19~23~5.37
+4~56~19.79 (quoted uncertainty $\pm$0.2~\arcsec).

On day~187, the VLA detected the nova at 3.5 and 6.1~cm, with peak
fluxes and total flux density as given in Table~\ref{tab-VLA}.  At
20~cm confusing sources make the results difficult to assess, but
estimated upper limits on the peak flux and total flux density are
also given in Table~\ref{tab-VLA}. The emission was unresolved and
peak positions in Table~\ref{tab-VLA} were determined from Gaussian
fits. These are consistent with both the MERLIN and \citet{Kiss04}
positions, allowing for the relatively low resolution of the VLA in
C--array.

\begin{table}
\begin{tabular}{lllll}
Wavelength & Peak              & Total & RA$^*$ & Dec$^*$\\
(cm)      & (mJy~beam$^{-1}$) & (mJy) & (s)    & (arcsec)\\
\hline
20  &  19$\pm$2$^{**}$ & 20$\pm$2$^{**}$ & 5.390 & 20.19\\
6.1 & 33.6$\pm$0.2 & 35.4$\pm$0.2 & 5.322 & 19.71\\
3.5 & 56$\pm$4     & 59$\pm$4     & 5.337 & 19.59\\
\end{tabular}
\caption{Observed parameters of V1494~Aql on 2000~June~7, day~187,
from VLA observations. $^*$Peak positions relative to 19~23~00
+4~57~00 and are consistent with the optical position of
\citet{Kiss04}. $^{**}$These should be regarded as estimated upper
limits due to confusing sources.}
\label{tab-VLA}
\end{table}

\begin{figure*}
\includegraphics[width=17.7cm]{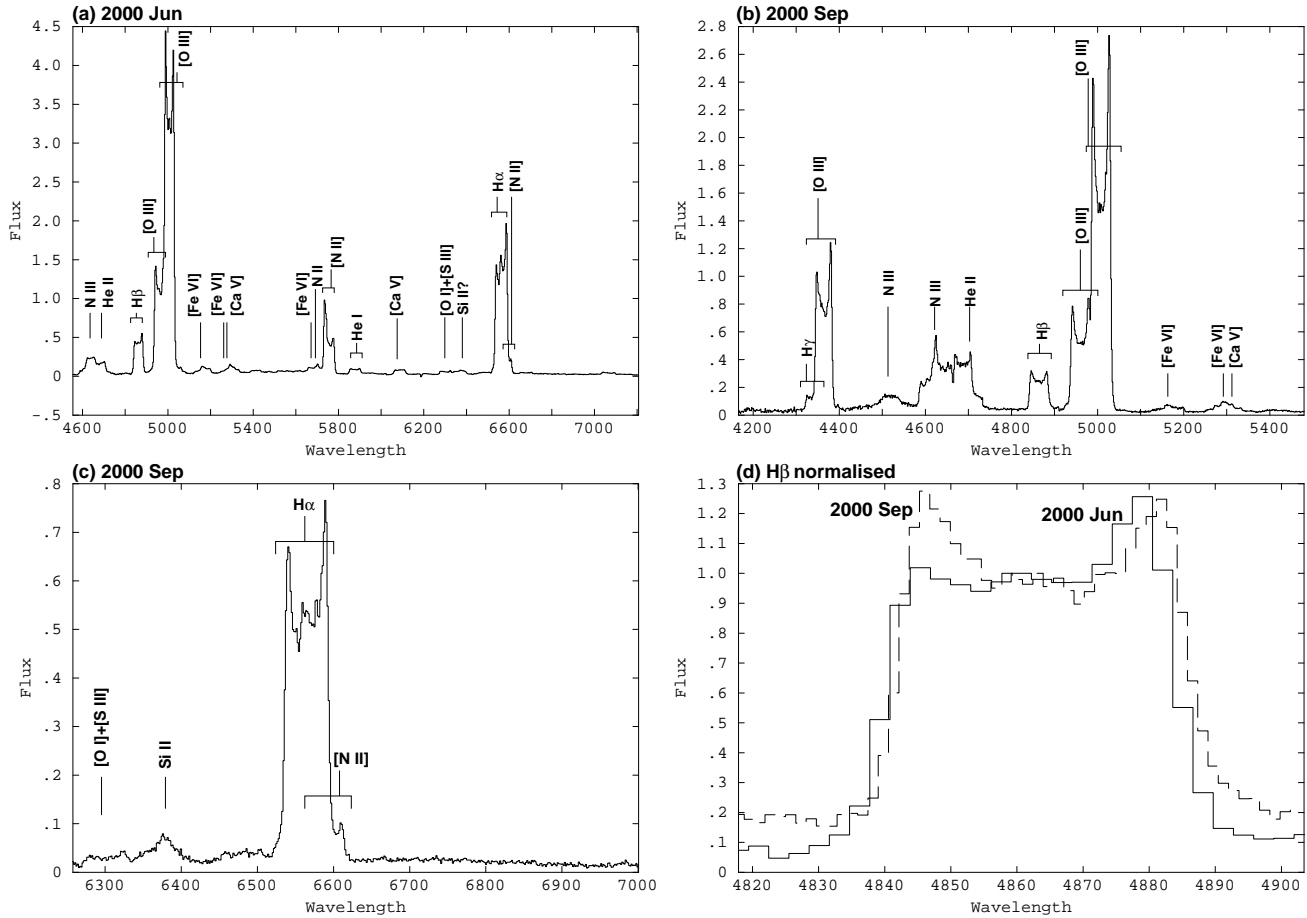}
\caption{(a)--(c) Optical spectra with line identifications as
indicated. They are not flux calibrated. (d) H$\beta$ profiles from both
dates, normalised to the central peak. The slight shift between the
peaks is instrumental.}
\label{fig-spectra}
\end{figure*}

Optical spectra were taken on days~194 and 195, and again on days~281
and 286.  Fig.~\ref{fig-spectra} shows these spectra, which clearly
exhibit emission from H$\alpha$, H$\beta$, H$\gamma$, He~\I,
[O~\I\I\I] and [N~\I\I]. 
In addition the blended lines possibly include He~\I\I, [Fe~\I\I\I],
[Fe~\V\I], [Fe~\V\I\I], [Ca~\V], C~\I\V\ and Si~\I\I, although many of
these are tentative identifications due to the breadth of the
features.  In all cases where individual species are reasonably
distinct or dominant, the line profiles are consistent with a complex
velocity structure in the ejecta.  This type of profile has been seen
in previous classical novae \citep{Gill99}, but it is apparent that
the width of the lines in V1494~Aql means that adjacent species are
blended in most cases.  In particular the [N~\I\I] $\lambda$6583.6\AA\
profile will have a blue peak merged with the H$\alpha$ profile. Only
H$\beta$ is sufficiently distinct to draw further conclusions.  In
general the profiles on the two dates show very similar peak
wavelengths for each component, symmetric around the rest
wavelength. In 2000~June the blue--shifted component appears to be
suffering absorption with respect to the red--shifted component; by
2000~September the two peaks are of comparable strength consistent
with a reduction in the optical depth at this wavelength.

We have carried out 5--component Gaussian fits to the H$\beta$
profiles in order to find the wavelengths of outlying peaks.  These
values, along with estimated Full Width Half Maximum (FWHM) and Full
Width Zero Intensity (FWZI) for each date, are given in
Table~\ref{tab-Hbeta}.  There is evidence for acceleration in the
early decline \citep{Iijima03} but none here for broadening of the
profiles with time. The FWHM of the lines are consistent with values
around 2700~km~s$^{-1}$ in 2000~June and September presented by
\citet{Iijima03}. However there is no evidence for the variable
high-velocity components that they found in their spectra of
2000~March. 

\begin{table*}
\begin{tabular}{lcccccc}
Date           & \multicolumn{2}{c}{Peak separation} &
               \multicolumn{2}{c}{Full Width Half Maximum} &
               \multicolumn{2}{c}{Full Width Zero Intensity}\\
               & \AA & km~s$^{-1}$ & \AA & km~s$^{-1}$ & \AA & km~s$^{-1}$ \\
\hline
2000 June      & 35.7$\pm$2.4 & 2200$\pm$150 & 45.3$\pm$6.0 &
               2790$\pm$370& 61$\pm$4 & 3760$\pm$250\\
2000 September & 37.7$\pm$1.0 & 2320$\pm$60 & 44.8$\pm$3.1 &
               2760$\pm$190 & 66$\pm$4 & 4070$\pm$250\\
\end{tabular}
\caption{Separation of the outer--most peaks, FWHM and FWZI for the
H$\beta$ profiles in Fig~\ref{fig-spectra} by wavelength and
velocity equivalent.  Determined from a
5--component Gaussian fit in both cases.}
\label{tab-Hbeta}
\label{tab-velocities}
\end{table*}

\section{Discussion}
\label{sec-discussion}

\subsection{Model fitting}
\label{ssec-model_fitting}

Models for radio emission from novae have generally assumed it
originates in free--free emission from an expanding isothermal shell.
Initially the shell is optically thick so that the flux density is
proportional to frequency squared ($\nu^2$) at any instant and rises
as time squared ($t^2$) as the shell expands. Eventually the shell
becomes optically thin and the flux density becomes proportional to
$\nu^{-0.1}$ at any given instant.  The flux density then declines as
the shell becomes more dilute \citep[see][for further
discussion]{Hjellming96}.

It is worth considering how the cm and sub--mm fluxes seen in the
outburst of V1494~Aql compare to the predictions of such models. At
the earliest epoch (4--5~d after outburst) the sub--mm observations
give a spectral index $\alpha$=2.5$\pm$0.5 ($F_{\nu} \propto
\nu^{\alpha}$), which is consistent with an optically thick
shell. With $\alpha$=2 we estimate that the flux density at 6.1~cm at
the same time would be $\sim4.4\times10^{-3}$~mJy.  We do not have an
observation at 6.1~cm at this time.  However, we can use the $t^2$
prediction of the simple models to extrapolate the estimate at 5~d and
predict a 6.1--cm flux density of $\sim6.2$~mJy by day~187. This falls
well below the observed value of 35.4~mJy. The spectral index at this
time between 6.1~cm and 3.5~cm is $\sim$1, suggesting that the shell
was beginning to become optically thin. If that was the case, the
extrapolated flux density of 6.2~mJy would represent an upper
limit. Similarly extrapolating from day~5 to day~27, assuming the
emission remains optically thick, leads to a prediction for the flux
density at 18~cm of around 0.02~mJy, a factor 25 lower than that
detected by MERLIN. The conclusion is that a single simple isothermal
model cannot be used to simultaneously fit the early sub--mm
observations and the later cm observations. This is in agreement with
the findings of \citet{Ivison93} for the case of V1974~Cyg.

Finally the flux density at 6~cm on day~137, from MERLIN observations,
compared with the VLA flux density at 6.1~cm on day 188, suggests a
rise at $\sim t^{1.1}$. This is consistent with the simple model at
turn over between the optically thin and optically thick
regimes. However by this stage MERLIN may well be resolving out some
of the structure at 6~cm and hence this flux density must be
considered as a lower limit. We note that changes to the H$\beta$
profiles between 2000~June and 2000~September (days 195 and 282; see
Fig.~\ref{fig-spectra}) are consistent with the gas becoming optically
thinner at optical wavelengths over that period.

\begin{figure}
\includegraphics[angle=0, width=8.5cm]{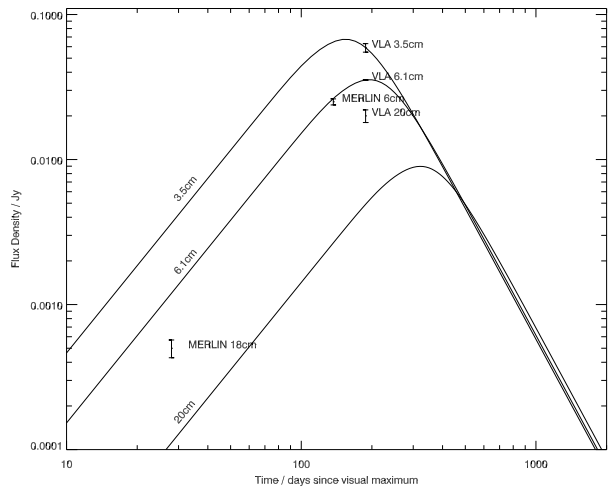}

\caption{Radio light--curves from the Hubble--flow model, fitted to
the VLA data and MERLIN data as marked. The VLA 20~cm data point
should be considered an upper limit (see Table \ref{tab-VLA}).}
\label{fig-models}
\end{figure}

We have attempted to fit the data with the two standard models for
cm-wave radio emission from novae, the Hubble--flow model
\citep{Hjellming96} and the variable--wind model \citep{Kwok83}. As
our analysis above indicates it is impossible to obtain a satisfactory
fit to the whole submm--cm dataset. Of the two models, the
Hubble--flow option provides the results most consistent with the
data, considering that the VLA 20~cm flux is an upper--limit. The
results of our best fit to the VLA data and the MERLIN data are given
in Fig.~\ref{fig-models}.  From this we find ejection velocities of
$\sim4600$~km~s$^{-1}$ and a total shell mass of
2.8$\times$10$^{-4}$~M$_\odot$. While the velocity is high even for a
nova as fast as this one, the mass is typical of those determined from
this kind of observation \citep{Hjellming96}. See \citet{Heywood04}
for a discussion of these models and our application of them to
classical novae.

\subsection{Oxygen line ratios}
\label{ssec-oxygen_ratios}

In 2000~September we observed the three [O~\I\I\I] lines at
$\lambda\lambda$4363.5, 4958.9 and 5006.8\AA.  Following
\citet{Iijima03} and using their equation~1 for the intensity ratio
between the sum of the intensities of the two lines around 5000\AA\ to
that of the line at 4363\AA\ we can determine the electron density
n$_e$, using a nebula temperature of 10,700~K \citep{Iijima03}.  From
Fig.~\ref{fig-spectra}(b) we find a ratio of $3.33\pm0.77$.  This
gives a global average
n$_e=(1.15\pm0.37)\times10^{12}$~m$^{-3}$. Taking a rough ratio of the
summed profiles around 5000\AA\ to the profile at 4363\AA\ shows that
this ratio is fairly constant across the range of velocities,
suggesting little variation in density across the velocity structure
(and hence presumably across the spatial structure). This value is
comparable to but significantly smaller than the value of
5.2$\times10^{12}$~m$^{-3}$ from \citet{Iijima03} around the same
date.

\subsection{Remnant structure}
\label{ssec-remnant_structure}

In Fig.~\ref{fig-MERLIN} we can see that the remnant consists of a
partial shell, with two areas of extended emission approximately to
the east and west of the central binary system. The structure is
consistent with that apparent in observations of previous classical
novae \citep{Eyres96, Eyres00, Heywood04}. Note that the registration
between the optical and radio positions is not accurate enough to
allow us to place the star precisely with respect to the radio
structure. We assume that the star lies between the two radio peaks,
i.e. at the centre of the partial shell.

The spectra in Fig.~\ref{fig-spectra} are predominantly
double--peaked, consistent with emission from a shell of material.
\citet{Gill99} have demonstrated that multiply--peaked spectra are
consistent with sub--structure in the shell.  However, in the case of
V1494~Aql the lines are so broad that adjacent features merge, so that
it is impossible to distinguish sub--structure in one line from the
overall structure in an adjacent line.  Indeed where lines are most
likely isolated (e.g. in the case of H$\beta$) the profile has little
consistent sub--structure, and is essentially symmetric between the
two wings.  Table~\ref{tab-Hbeta} gives the calculated corresponding
expansion velocities for a simple shell model. 

\citet{Anupama01} present spectra taken between days~3 and 22.  They
show dramatic changes in the H$\alpha$ profile, from a P~Cygni profile
on day~3, to a broad, structureless form on day~5 \citep[also seen
by][]{Kiss00}, followed by the development of structure similar to
that seen in Fig.~\ref{fig-spectra} by day~22. It is interesting that
our JCMT observations coincide with the featureless stage of the
H$\alpha$ profile, suggesting that the ejecta were in transition from
optically thick to optically thin.  It is also of note that the
overall structure of the H$\alpha$ line profile remains similar from
day~15 right through to day~195, before the blue--shifted component
brightens by day~285. \citet{Hjellming96} attributed this switch from
a flat, broad spectrum to the form seen in Fig.~\ref{fig-spectra}(d)
to a transition from continuous mass outflow to a detached shell. This
would indicate a chronology in which continuous mass outflow ceases
prior to day~15, followed by the expansion of the optically thick
detached shell. However we are not convinced that the changes in the
profile require the end of mass outflow. For example a equatorial ring
of enhanced density would give a double--peaked profile even while
mass outflow continued. We note that \citet{Iijima03} claim evidence for
continued mass ejection more than 150~days after outburst. In either
case, the ejecta becomes optically thin in the visible wavelength
range between days~195 and 285. This places some time constraints on
future modelling of nova ejecta.

A simple Gaussian fit to the radio emission structure indicates that
on day~137 the major axis is 265$\pm$16~mas at position angle (east of
north) 110.7$\pm$4.2~degrees, minor axis 155$\pm$9~mas.  This agrees
well with the extent of the emission on the image
(Fig.~\ref{fig-MERLIN}).  In addition the separation of the peaks of
the two radio components is 120$\pm$5~mas. The position angle differs
from that of $\sim$140~degrees found around optical maximum by
\citet{Kawabata01} using spectropolarimetry. However given the
breadth of the radio features in Fig.~\ref{fig-MERLIN} the image seems
consistent with extension in the general south--east to north--west
direction, as seen at early times by \citet{Kawabata01}.

\citet{Kiss00} use a number of Maximum--Magnitude--Rate--of--Decline
relationships to find a distance of 3.6$\pm$0.3~kpc.  \citet{Iijima03}
find reddening amounting to A$_{\rm V}$ = 1.8, and consequently a
modified distance of 1.6$\pm$0.2~kpc. Using this latter distance
estimate we can use the observed extent of the radio emission to
estimate an ejection velocity. This leads to values ranging from
980$\pm$80~km~s$^{-1}$ to 2500$\pm$200~km~s$^{-1}$ depending on
whether the major or minor axes or the peak to peak separation are
used in the calculation. This velocity range is entirely consistent
with that derived from the spectra and presented in
Table~\ref{tab-Hbeta} and confirms that the radio continuum arises
from the same material as the optical lines. Thus it may be that at
this stage of the development of the radio emission (whilst the
remnant is still largely optically thick at these wavelengths) it is
valid to compare the extent of the remnant on the sky
(Fig.~\ref{fig-MERLIN}) with velocities derived from hydrogen spectra
to derive a distance to classical novae.

It has been suggested \citep{Kiss00} that the spectral line profiles
are consistent with a nearly edge--on equatorial ring of the sort
modelled by \citet{Gill99}.  We feel given the width of the individual
components this conclusion cannot be directly supported; we do not
rule out such a structure but the spectra do not require
it. Nonetheless we can examine the possibility that our MERLIN image
(Fig.~\ref{fig-MERLIN}) represents emission from a tilted
ring. Assuming that the major axis of the brightness distribution on
the sky represents the diameter of an equatorial ring, we take
foreshortening due to the tilt to lead to the
ellipticity. Consequently the minor axis can be compared with the
major axis to give an angle of inclination to the line--of--sight of
$\sim36.0^\circ\pm4.0^\circ$, which cannot be considered
edge--on. However it does agree well with the findings of
\citet{Iijima03} who reconcile high radial velocities of absorption
lines in the early decline with lower emission line widths in the
later nebular stage. They suggest the ejecta take the form of a ring
with an angle to the line of sight of 30$^\circ$, also in agreement
with our angle above. In this case the intrinsic velocities are twice
as large as those observed and may at least partly explain why the
high velocities derived from the spherically symmetric model for the
radio emission presented in section~\ref{ssec-model_fitting} are
inconsistent with the velocities derived from the optical spectroscopy.

We note \citet{Iijima03} also suggest the presence of high--velocity
jets in V1494~Aql.  These could also be a source of the elongation in
the radio emission, and the velocities of the jets are consistent with
emission of the angular extent seen. However it is difficult to see
how the radio emission could be reconciled with both fast jets and an
inclined ring of material; either one or the other must have caused
the observed structure. Given that our MERLIN observations are
contemporaneous with the Nebular stage as defined by \citet{Iijima03}
it seems simplest to associate the observed structure in the radio
with an inclined ring; the size scale also agrees with this
interpretation. This does not mean jets were not present also, only
that they were not visible to MERLIN.

\subsection{Remnant mass}
\label{ssec-remnant_mass}

We can use some very simple assumptions to gain a model--independent
estimate of the mass of the ejecta. Assuming ejecta formed of pure,
ionised hydrogen, the mass might be estimated as
\begin{equation}
m(H) = n_e m_H V ~{\rm kg}
\end{equation}
for an electron density $n_e$, hydrogen atom mass $m_H$ and ejecta
volume $V$. We estimate $V$ by assuming that the major and minor axes
of the emission on the sky are the major and minor axes of the
ellipsoidal ejecta, and that the intermediate axis is the mean of
these two. Taking into account continous mass ejection, we find the
volume of the shell occupied by the ejecta on day~136 to be
$V\sim6.3\times10^{40}$~m$^3$ at a distance of 1.6~kpc. This gives
$m(H)\sim6\times10^{-5}$~M$_\odot$. This is comparable with the total
mass estimate of \citet{Iijima03}, but significantly less than our
estimate from model fitting. If mass ejection had ended on day~15, the
ejecta would only occupy around 30\% of this volume, with a
corresponding $m(H)\sim1.8\times10^{-5}$~M$_\odot$.

\section{Conclusions}
\label{sec-conclusions}

We present a number of observations of the remnant of the classical
nova V1494~Aql.  These begin 4~d after visual maximum with sub--mm
data from the JCMT, followed by cm--band data after 136~d (MERLIN) and
187~d (VLA) and finally optical spectroscopy on $\sim$194~d and
$\sim$284~d (UAGS).  Together these observations probe the ejecta over
a period of 20~weeks, from the early ``fireball'' stage to the final
optically thin phase as the radio emission begins to fade away.
Comparison between the radio image and spectroscopy is consistent with
a distance of 1.6~kpc.

Both the optical spectra and the radio imaging demonstrate the early
clumpy structure of the ejecta around classical novae.  Most
importantly it conclusively shows that simple spherical models of the
ejecta development are insufficient to predict the behaviour over the
time and wavelength domain sampled here (183~days, 450$\umu$m to
20~cm).  Clearly new models for radio emission from novae must be
developed which address departures from spherical symmetry and the
clumpy nature of the ejecta. Furthermore it is essential to
investigate in detail how the predicted brightness distributions from
such models would be seen by radio interferometers such as MERLIN if
we are to confidently reconcile features in the radio images with
structures in the ejecta. We find that the emission is consistent with
either an inclined ring or jets of ejecta \citep[as suggested
by][]{Iijima03} but not both; we are inclined to associate it with the
former rather than the latter.  We also find data consistent with mass
ejection continuing for six months after the outburst right up to the
ejecta starting to become optically thin in the visible regime after
day 195. This continuous mass ejection needs to be tested in other
novae, in particular looking for correlations with speed class.

\section*{Acknowledgments}

\noindent VE and IH acknowledge the support of PPARC. The JCMT is
operated by the Joint Astronomy Centre in Hilo, Hawaii on behalf of
the parent organizations Particle Physics and Astronomy Research
Council in the United Kingdom, the National Research Council of Canada
and the Netherlands Organization for Scientific Research. MERLIN is a
National Facility operated by the University of Manchester at Jodrell
Bank Observatory on behalf of PPARC.  The National Radio Astronomy
Observatory is a facility of the National Science Foundation operated
under cooperative agreement by Associated Universities, Inc. This work
used the on--line service of VSNET.

\bsp

\label{lastpage}
\end{document}